\documentclass[12pt]{article}  

\usepackage{graphicx,amsmath}
\usepackage{units}

\newlength{\dinwidth}
\newlength{\dinmargin}
\renewcommand{\vec}[1]{\boldsymbol{#1}}

\def\lapproxeq{\lower .7ex\hbox{$\;\stackrel{\textstyle                                                    
<}{\sim}\;$}}                                                    
\def\gapproxeq{\lower .7ex\hbox{$\;\stackrel{\textstyle                                                    
>}{\sim}\;$}}                                                    
\def\be{\begin{equation}}                                                    
\def\ee{\end{equation}}                                                    
\def\bea{\begin{eqnarray}}                                                    
\def\eea{\end{eqnarray}}

\def\sh{\hat s}
\def\sh2{{\hat s}^2}
\begin{document}                                                    
\titlepage                                                    
\begin{flushright}                                                    
IPPP/10/85  \\
DCPT/10/170 \\                                                    
\today \\                                                    
\end{flushright} 
\vspace*{0.5cm}
\begin{center}                                                    
{\Large \bf A note on rapidity distributions at the LHC }

\vspace*{1cm}
                                                   
V.A. Schegelsky$^{a}$, M.G. Ryskin$^{a,b}$, A.D. Martin$^b$ and V.A. Khoze$^{a,b}$ \\                                                    
                                                   
\vspace*{0.5cm}                                                    
$^a$ Petersburg Nuclear Physics Institute, Gatchina, St.~Petersburg, 188300, Russia \\            
$^b$ Institute for Particle Physics Phenomenology, University of Durham, Durham, DH1 3LE \\

\vspace*{1cm}                                                    
                                                    
\begin{abstract}                                                    

We discuss the difference between the distribution of secondaries measured in terms of pseudorapidity and that using the correct rapidity variable. We show a set of examples obtained using Monte Carlo simulations. We also consider the production of particles of low transverse momentum where coherence effects may occur, which are not yet included in the present Monte Carlos.
\end{abstract}                                                        
\vspace*{0.5cm}                                                    
                                                    
\end{center}                                                    
                                                    

\vspace*{0.3cm}

The purpose of this note is to recall the difference between rapidity. $y$, and pseudorapidity, $\eta$. This is not a new topic \cite{RPP}. However, since the results of the LHC detectors are shown mainly in terms of pseudorapidity, it is timely to discuss effects which may be caused by the identification of $\eta$ with $y$.

\section{Definitions of rapidity and pseudorapidity}
 The Lorentz-invariant single-particle inclusive cross section has the form
\be
f({\rm AB} \to {\rm CX})~\equiv~E\frac{d^3\sigma}{d^3p}~=~E\frac{d^3\sigma}{\pi dp_Ldp_T^2}
\ee
where $E,~p_L$ and $p_T$ are the energy, longitudinal and transverse momentum of the outgoing particle C with respect to the incoming proton direction; and $d^3p/2E$ is the Lorentz-invariant phase space
\be
\frac{d^3p}{2E}~=~d^4p~\delta(p^2-m^2)~\Theta (E),
\ee
where $m$ is the mass of particle C. For high-energy hadronic collisions, experiment shows that the longitudinal momentum may take almost any value, whereas the transverse momentum is usually small.

The rapidity $y$ is a convenient way to display the $p_L$ dependence of the data, where
\be
y~=~ \frac{1}{2} {\rm log}\left(\frac{E+p_L}{E-p_L}\right)~=~{\rm log}\left(\frac{E+p_L}{m_T}\right),
\ee
where $m_T^2=m^2+p_T^2$. The value of $y$ depends on the reference frame, and, at first sight, it looks like a rather ungainly construct.  However, there are advantages to present high energy data as a function of rapidity.  First, we see $dy=dp_L/E$, and so the Lorentz-invariant cross section
\be
f({\rm AB} \to {\rm CX})~=~\frac{d^3\sigma}{\pi dy dp_T^2}.
\ee
Second, rapidity is additive under Lorentz boosts along the beam direction. For example, under a boost of velocity $u$
\be
E \to \gamma (E+up_L),~~~~~~~p_L \to \gamma (p_L+uE),
\ee
where $\gamma=1/\sqrt{1-u^2}$. Hence
\be
y~\to~y~+~\frac{1}{2} {\rm log}\left(\frac{1+u}{1-u}\right)~=~y+y_{\rm boost}.
\ee
Thus the difference in rapidity of two particles is not changed by boosts along the beam axis.
In the non-relativistic limit, when the boost $v\ll 1$, we have $E\to m,~p\to~mv$ and thus the rapidity $y~\to$ velocity $v$ (hence its name).

For a massless particle $E=|\vec{p}|$.  Then the rapidity can be directly expressed in terms of the scattering angle $\theta$
\be
y~=~\frac{1}{2} {\rm log}\left( \frac{1+{\rm cos}\theta}{1-{\rm cos}\theta}\right)~=~-{\rm log ~tan}\frac{\theta}{2}.
\ee
In general, we define pseudorapidity as
\be
\eta~=~-{\rm log ~tan}\frac{\theta}{2}.
\ee
Clearly, for massless particles the rapidity and pseudorapidity are the same. However, this is also the case for particles with $p_T \gg m$ and $p_L \gg m$.

\begin{figure} 
\begin{center}
\includegraphics[height=5cm]{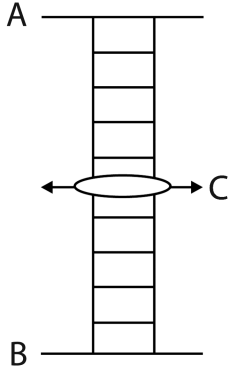}
\caption{\sf A sketch showing the inclusive production of particle C.}
\label{fig:A1}
\end{center}
\end{figure}

\section{Expectations at the LHC}
\begin{figure} 
\begin{center}
\includegraphics[height=8cm]{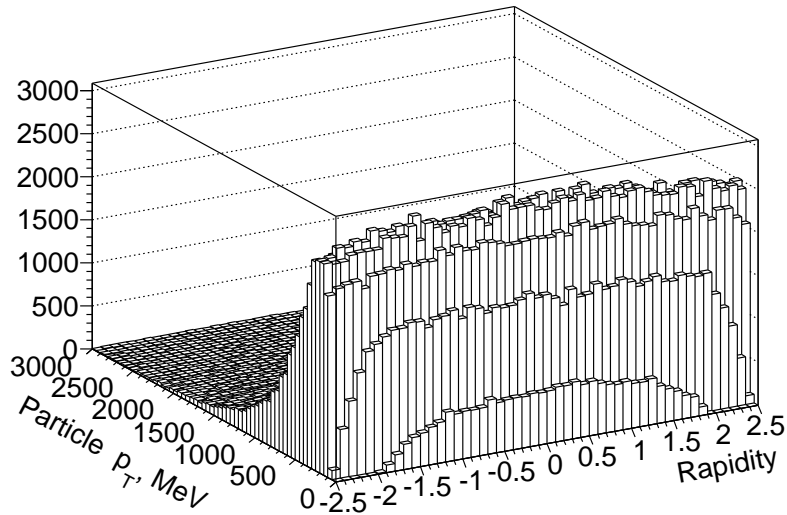}
\caption{\sf The $p_T$-rapidity correlation obtained for pions from a Monte Carlo sample.}
\label{fig:A2}
\end{center}
\end{figure}

We now come to the single-particle rapidity distributions measured at the LHC.  From the theoretical viewpoint, both the phase space of particle C and the matrix element should be written in terms of covariant variables, that is in terms of $y$ and $p_T$.  Long ago, Regge theory predicted a flat rapidity distribution in the central region. Indeed, in Regge theory it assumed that $p_T$ is limited and $\langle p_T \rangle$ does not increase with energy. At very high energies, the inclusive cross section is described by the emission of secondaries from the Pomeron. The probability of the emission of particle C is given by the effective Pomeron-C vertex, $\beta_{PC}$, which does not depend on rapidity \cite{Mueller,Kancheli}, see Fig.~\ref{fig:A1}. We have
\be
\frac{d\sigma}{\pi dy dp_T^2}~=~\beta_N^2~\beta_{PC}(p_T)~s_{AC}^{\alpha (0)-1}~s_{BC}^{\alpha (0)-1}.
\ee
If $y$ is measured in the $pp$ centre-of-mass frame, then
\be
s_{AC,BC}~=~m_p~m_T~ {\rm exp}(\pm y+Y/2)
\ee
(where $s_{AC}=(p_A+p_C)^2,~Y={\rm ln}(s/m_p^2)$ and $\alpha(0)$ is the intercept of the Pomeron trajectory), and the independence of the distribution on $y$ is explicit. The cross section may also be written in terms of parton distributions
\be
\frac{d\sigma}{ dy dp_T^2}~=~\int dx_1 dx_2~ g(x_1)  g(x_2)~\frac{d\sigma(gg \to C)}{dp_T^2}~\delta\left(y-\frac{1}{2}{\rm ln}\frac{x_1}{x_2}\right),
\label{eq:parton}
\ee
with $x_{1,2}=(m_T/\sqrt{s}){\rm exp}(\pm y)$. If the gluon distributions can be described by a fixed power, $g(x) \sim x^{-\lambda}$, then again we see that the single-particle multiplicity distribution is expected to be flat in $y$. Indeed, the present Monte Carlos are based on (\ref{eq:parton}), and yield a flat distribution in the central region, see Fig.~\ref{fig:A2}. The PYTHIA 6.4 Monte Carlo (ATLAS) \cite{Pythia} was used to demonstrate the effects in Figs.~\ref{fig:A2}$-$\ref{fig:A7}.

\section{The effects of approximating $y$ by $\eta$}
\begin{figure} 
\begin{center}
\includegraphics[height=8cm]{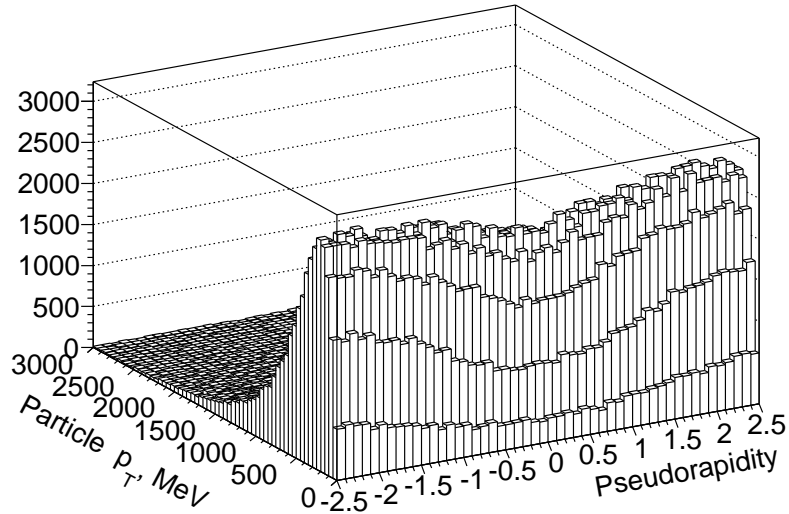}
\caption{\sf The $p_T$-pseudorapidity correlation obtained from the same Monte Carlo sample as Fig~\ref{fig:A2}.}
\label{fig:A3}
\end{center}
\end{figure}

 The single-particle distributions so far measured at the LHC have limited particle identification. The three-momentum of the particle is measured, but not, in general, its mass. Without knowledge of $m$ it is not possible to calculate $y$. Therefore the results are presented in terms of the pseudorapidity $\eta$.  For $\eta=y$ we need $p_T \gg m$. However,  for low $p_T$ the equality does not hold. In particular, a particle with low $p_T$ may have small $y$, but rather large $\eta$, since for low $p_T$ this slow particle tends to be emitted at a small polar angle $\theta$. For simplicity, we consider $p_L \gg m$. Then the difference
\be
\eta-y~=~{\rm log}\frac{m_T}{p_T}~+~O\left(\frac{m_T^2}{p_L^2}\right).
\ee
The log may be quite large as $p_T \to 0$. The most interesting region, $\eta \sim 0$, in the multiplicity distributions measured at the collider, contains a singularity\footnote{For fixed $p_T$, the Jacobian between the $dy$ and $d\eta$ bins is $d\eta =(E/|\vec{p}|)~dy$. For low $p_T$, it has a singularity at $|\vec{p}| \to 0$.} at $\eta=0$. As a consequence the low $p_T$ particles are spread over a larger $\eta$ region, depopulating the $\eta \sim 0$ region. This is clearly seen in Fig.~\ref{fig:A3} as a dip in the $\eta$ distribution for $\eta \sim 0$. For distributions corresponding to larger values of $p_T$ the dip is less pronounced.

\begin{figure} 
\begin{center}
\includegraphics[height=8cm]{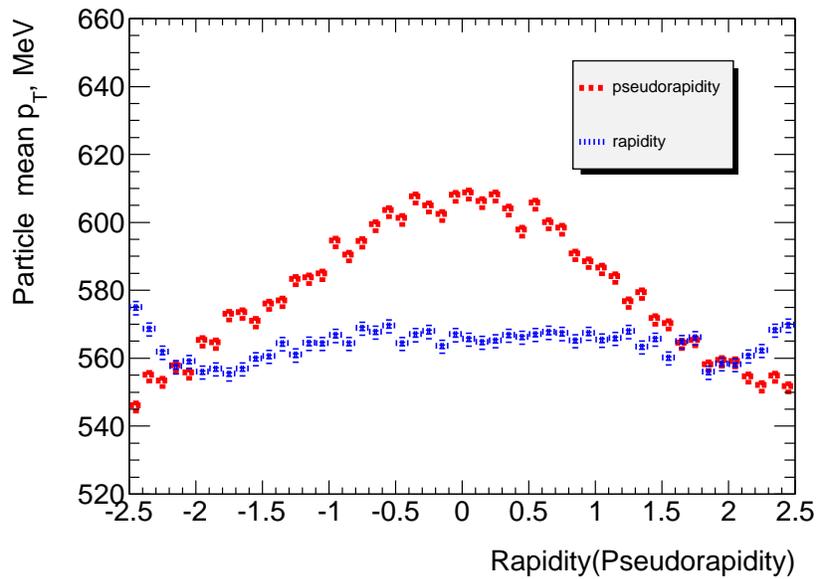}
\caption{\sf The mean value of $p_T$ as a function of pseudorapidity (upper points) and rapidity (lower points).}
\label{fig:A4}
\end{center}
\end{figure}

This trivial kinematic effect produces some $\eta$ dependence of $\langle p_T \rangle$ measured in the central detector. Since low $p_T$ particles are spread over a larger number of low $\eta$ bins, the measured value of $\langle p_T \rangle$ decreases with increasing $|\eta|$, see Fig.~\ref{fig:A4}. This effect is not present in the $y$ distributions.

\begin{figure} 
\begin{center}
\includegraphics[height=8cm]{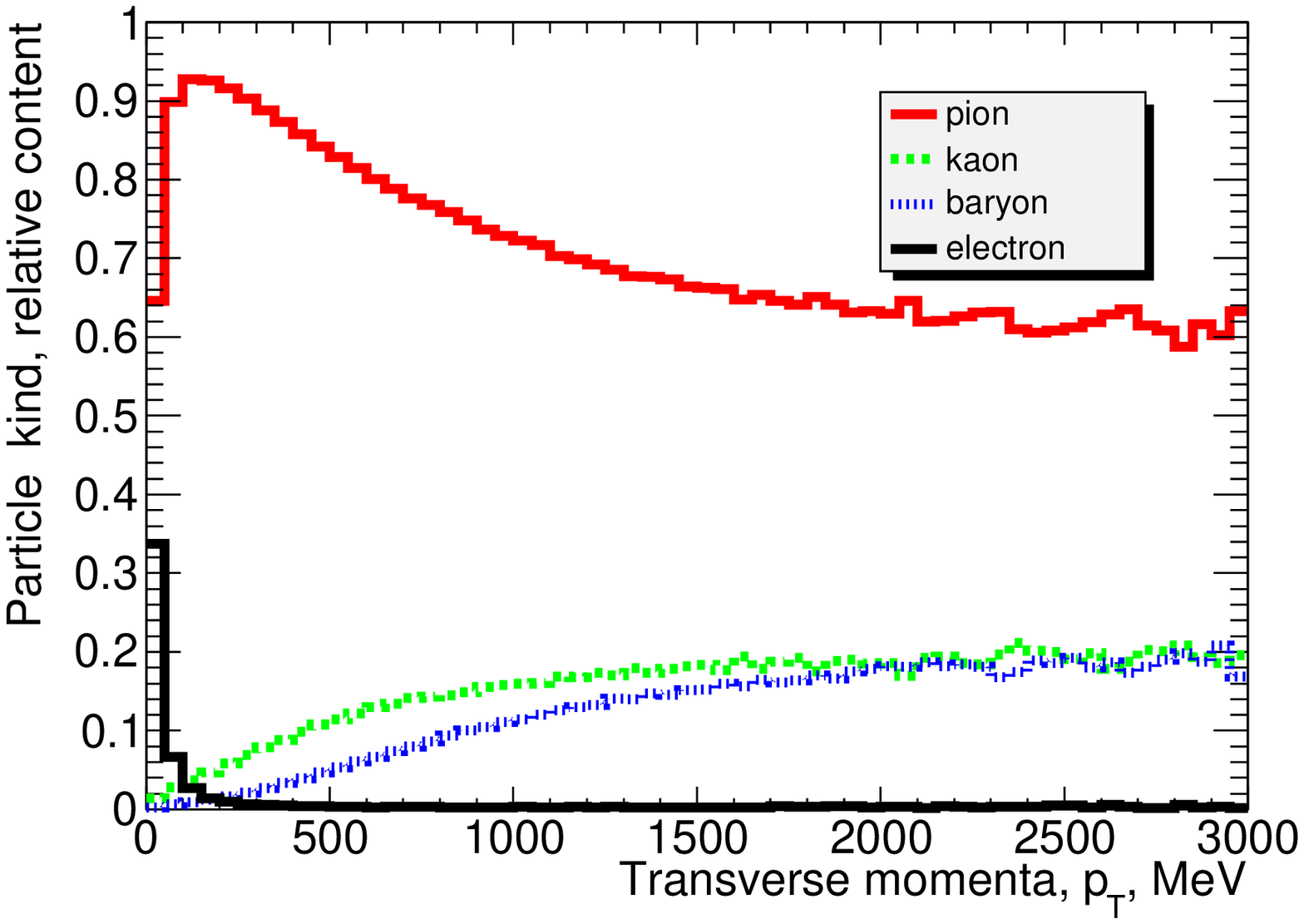}
\caption{\sf The Monte Carlo prediction for different types of minimum bias particles as a function of their $p_T$. Although the pions dominate, there are a appreciable number of soft electrons in the very lowest $p_T$ bins and a significant kaon and baryon component at the larger $p_T$ values.}
\label{fig:A5}
\end{center}
\end{figure}
\begin{figure} 
\begin{center}
\includegraphics[height=8cm]{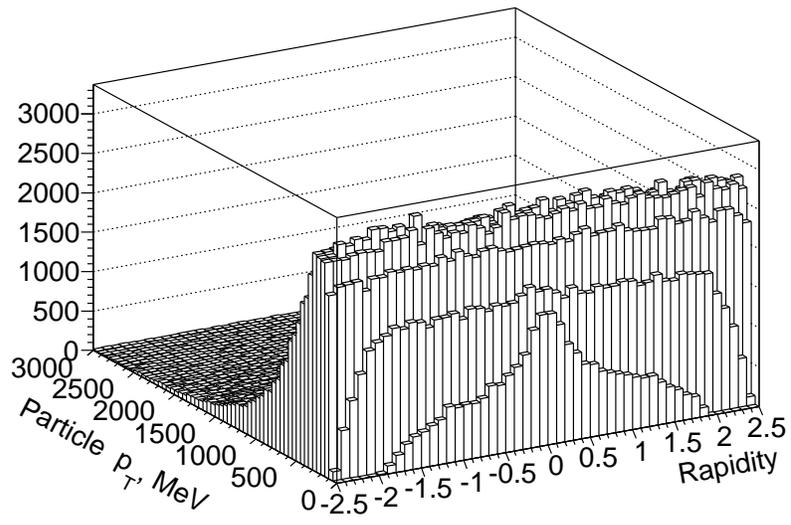}
\caption{\sf The $p_T$-rapidity correlation assuming that all the particles of Fig.~\ref{fig:A5} have the mass of the pion. }
\label{fig:A6}
\end{center}
\end{figure}
\begin{figure} 
\begin{center}
\includegraphics[height=8cm]{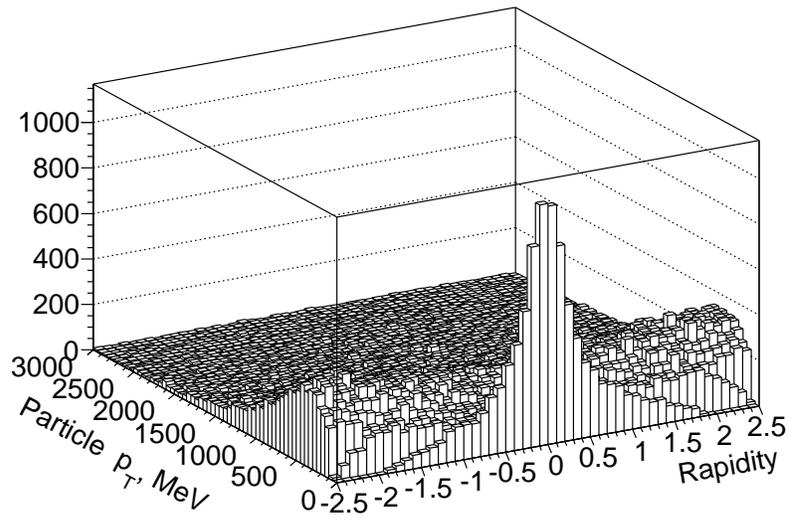}
\caption{\sf The same as Fig.~\ref{fig:A6}, but omitting the pion component and taking the other particles to have the mass of the pion.}
\label{fig:A7}
\end{center}
\end{figure}

Of course, the majority of secondaries are pions. Therefore the multiplicity distributions may be presented in terms of $y$ by assigning all particles to be pions. Nevertheless there are two sources of potential difficulty. First look at Fig.~\ref{fig:A5}. It shows the fractions of different sorts of secondaries according to a PYTHIA Monte Carlo simulation. First, we see the presence of electrons (arising from Dalitz pairs) in the very lowest $p_T$ bins. If the electron is assigned the mass of the pion, then its `effective' rapidity becomes close to zero, which produces a prominant erroneous peak for $y \sim 0$. Second, at larger $p_T$, the kaons and baryons, which are misassigned as pions, lead to an erroneous dip for $y \sim 0$. This dip disappears as we go to larger values of $p_T$. Note however, that to neglect the difference between $y$ and $\eta$, we need a value of $p_T$ larger than the mass of the corresponding hadron. All these effects are seen in Figs.~\ref{fig:A6} and \ref{fig:A7}.  Note, from Fig.~\ref{fig:A5}, that in the larger $p_T$ region, the fraction of kaons and baryons is appreciable. Only about 60$\%$ of secondaries are pions.
Recall also that the region of low $p_T$ and small $\eta$ is very poorly described by the present Monte Carlo models. We should therefore take care when discussing this domain.

\section{Examples of the necessity to use $y$ rather than $\eta$}
To demonstrate the importance of using the correct rapidity variable, we consider the coherence effects in low $p_T$ particle production.
In the Regge approach, the $p_T$ distribution of secondaries  produced in the central region ($y \simeq 0$) does not depend on energy,
but reflects only the internal $p_T$-structure of the Pomeron.
Therefore we expect that the particle density $dN/dy d p^2_T$ measured at low $p_T$ will increase with energy just like the density $dN/dy$ integrated over\footnote{The region of a large $p_T$ gives a small contribution.} over $p_T$.  That is, by about 60\%
in going from the collider energy of 0.9 to 7 TeV. In terms of Monte Carlo models such an effect is simulated by the number of Multiple Interactions (MI), which increases with initial energy of the colliding protons. 

On the other hand, such an approach does not account for the coherence effects which may be quite important for low momenta secondaries. 
For low $p_T$ particles, interference  arises when a particle is radiated from different lines of Feynman diagrams which make up the whole matrix element. One example of this is soft gluon emission. Another example is the Hanbury-Brown$-$Twiss effect, where two identical particles can be radiated from different places in the interaction region.  Their interference leads to a peak in the distribution of the two identical particles when their momenta are close to each other. The width of the peak can be used to estimate the size of the interaction region \cite{HBT}.

First we discuss soft gluon emission, which leads to the so-called limiting energy behaviour.  The radiation of soft gluons from several quarks and/or gluons develops in a coherent way. The reason is that  different sources of secondaries should act as a single source
 with an effective colour `charge' equal to the `{\it vector}' sum 
of all the colour `charges', as a gluon of large wavelength cannot resolve these smaller details. Recall that, due to the Lorentz time dilation, the low momentum hadrons are formed at the initial stage, before the  fast particles \cite{book}. These fast particles, which are formed later on, fly out of the interaction region and leave the slow hadrons intact\footnote{An analogy is a supernova explosion where the remnants of the star are left intact.}. Such a limiting behaviour of soft secondaries was first predicted, and observed, in $e^+e^-$ annihilation \cite{K1,K2}, and then extended to the $pp$ case \cite{OKR}.

Also, note that, if the incoming hadron represents a  
 quark-antiquark (meson) or quark-diquark (nucleon) 
system, then the $t$-channel
 multi-gluon exchange corresponds dominantely to a colour octet.
The same colour-octet exchange dominates the BFKL amplitude.
So even in the case of a large number of gluon exchanges (or a large number of MI), the low $p_T$ secondaries will be produced mainly
via radiation caused by colour-octet flow.

Thus, in the ideal case, we may expect a limiting energy behaviour of the soft particle density. That is, at very low $p_T$, the value of $dN/dy dp^2_T$ should not depend (or depend only weakly) on the initial energy, see Ref.~\cite{OKR}.
However, as was shown above, in order to measure the density of low-$p_T$ particles at small $y$, it is important to use the correct `rapidity' variable, and not just the angle (that is, $\eta$). Otherwise the expected `limiting' behaviour may be affected by an admixture of other  sorts  of particles, some lighter and some heavier than the pion.

The Hanbury-Brown$-$Twiss correlation can be viewed as the first step in the formation of condensates of bosons which arise from Bose-Einstein statistics. At high energies when the particle density increases, and the number of particles exceeds one in the elementary $\Delta x\Delta p$ cells of some domain of configuration space, there may be two sorts of such collective phenomena. One is of pure classical, thermodynamic origin, like elliptic flow in quark-gluon plasma. Another one is based on quantum mechanical interference. We consider the formation of some multiparticle coherent state (such as in superconductivity or in a disorientated chiral condensate \cite{DCC}). The coherent emission of soft gluons by the vector sum of the whole colour charge may be considered as the first signature of the multiparticle coherent state. Since the wavelength of low $p_T$ particles is large, it is expected that such collective phenomena will occur, and be best observed, in the low $p_T$ region. In comparison with lower energies, these collective, coherent effects should be enhanced at the LHC due to the larger density of low $p_T$ secondaries.

In summary, to observe such phenomena we need a careful study of very soft particle distributions using the correct Lorentz covariant variables.

\section*{Acknowledgements}

We thank Wolfgang Ochs for a fruitful discussion.

\end{document}